\documentclass[conference]{IEEEtran}
\IEEEoverridecommandlockouts
\usepackage{cite}
\usepackage{amsmath,amssymb,amsfonts}
\usepackage{amsmath,amsfonts}
\usepackage{algorithmic}
\usepackage{algorithm}
\usepackage{array}
\usepackage[caption=false]{subfig}
\usepackage{textcomp}
\usepackage{stfloats}
\usepackage{url}
\usepackage{verbatim}
\usepackage{graphicx}
\usepackage{tablefootnote}
\graphicspath{{images/}}
\usepackage{xcolor}
\usepackage{color}
\usepackage{marvosym}

\usepackage{lineno}
\usepackage{ulem}
\usepackage{float}
\usepackage{hyperref}
\usepackage{fancyhdr}

\makeatletter
\newcommand{\removelatexerror}{\let\@latex@error\@gobble}
\makeatother

\newcommand{\mbbE}{\mathbb{E}}
\newcommand{\mbbR}{\mathbb{R}}
\newcommand{\mbbC}{\mathbb{C}}

\newcommand{\mbB}{\mathbf{B}}

\newcommand{\mbG}{\mathbf{G}}
\newcommand{\mbH}{\mathbf{H}}
\newcommand{\mbI}{\mathbf{I}}
\newcommand{\mbL}{\mathbf{L}}
\newcommand{\mbQ}{\mathbf{Q}}
\newcommand{\mbR}{\mathbf{R}}

\newcommand{\mbT}{\mathbf{T}}
\newcommand{\mbDel}{\mathbf{\Delta}}
\newcommand{\mbPi}{\mathbf{\Pi}}
\newcommand{\mbXi}{\mathbf{\Xi}}

\newcommand{\mbSg}{\mathbf{\Sigma}}
\newcommand{\mbU}{\mathbf{U}}

\newcommand{\mcS}{\mathcal{S}}
\newcommand{\mcU}{\mathcal{U}}
\newcommand{\mcC}{\mathcal{C}}

\newtheorem{theorem}{\hspace{-1em}\textbf{Theorem}}

\newcommand{\abs}[1]{\left\vert#1\right\vert}
\newcommand{\norm}[1]{\left\Vert#1\right\Vert}
\newcommand{\sbrac}[1]{\left(#1\right)}
\newcommand{\lbrac}[1]{\left\{#1\right\}}

\def\BibTeX{{\rm B\kern-.05em{\sc i\kern-.025em b}\kern-.08em
    T\kern-.1667em\lower.7ex\hbox{E}\kern-.125emX}}
\begin{document}

\title{TOSE: A Fast Capacity Estimation Algorithm Based on Spike Approximations}


\author{\IEEEauthorblockN{Dandan Jiang\thanks{The author Dandan Jiang was supported by Key technologies for coordination and interoperation of power distribution service resource, Grant No. 2021YFB2401300, NSFC Grant No. 11971371 and the Fundamental Research Funds for the Central Universities.}}
	\IEEEauthorblockA{\textit{School of Mathematics and Statistics} \\
		\textit{ Xi'an Jiaotong University}\\
		Xi’an, China \\
		jiangdd@xjtu.edu.cn}
	\\
	\IEEEauthorblockN{ Lu Yang\textsuperscript{\Letter}}
	\IEEEauthorblockA{\textit{Theory Lab, Central Research Institute, 2012 Labs} \\
		\textit{Huawei Technology Co. Ltd.}\\
		Hong Kong, China \\
		yanglu87@huawei.com}
	\and
	\IEEEauthorblockN{ Han Hao}
	\IEEEauthorblockA{\textit{School of Aerospace Engineering} \\
		\textit{Tsinghua University}\\
		Beijing, China \\
		haoh19@mails.tsinghua.edu.cn}
	\\
	\IEEEauthorblockN{Rui Wang}
	\IEEEauthorblockA{\textit{School of Mathematics and Statistics} \\
		\textit{Xi'an Jiaotong University}\\
		Xi’an, China \\
		wangrui\_math@stu.xjtu.edu.cn}}
\maketitle

\vspace{-2em}
\begin{abstract}
	Capacity is one of the most important performance metrics for wireless communication networks.
	It describes the maximum rate at which the information can be transmitted of a wireless communication system.
	To support the growing demand for wireless traffic, wireless networks are becoming more dense and  complicated, leading to a higher difficulty to derive the capacity. Unfortunately,  most existing methods for the capacity calculation take a polynomial time complexity. This will become unaffordable for future ultra-dense networks, where  both the number of base stations (BSs) and the number of users are extremely large.
	In this paper, we propose a fast algorithm TOSE to estimate the capacity for ultra-dense wireless networks.  Based on the spiked model of random matrix theory (RMT), our algorithm can avoid the exact eigenvalue derivations of large dimensional matrices, which are complicated and inevitable in conventional capacity calculation methods. 
	Instead, fast eigenvalue estimations can be realized based on the spike approximations in our TOSE algorithm. 
	Our simulation results show that TOSE is an accurate and fast capacity approximation algorithm. Its estimation error is below 5\%, and it runs in linear time, which is much lower than the polynomial time complexity of existing methods. 
	In addition, TOSE has superior generality, since it is independent of the distributions of BSs and users, and the shape of network areas.
\end{abstract}


\section{Introduction}\label{section1}
Capacity can be regarded as one of the most important performance metrics of wireless communication networks.
Capacity determination for future wireless networks is also listed as one of the ten most challenging information and communication technology (ICT) problems in the post-Shannon era \cite{xu2021ten}.
With the rapid development of mobile communication technology,  wireless systems become more and more complicated, leading to a higher complexity in determining the capacity.
In 1948,  Dr. Claude E. Shannon defined the notion of channel capacity and provided a mathematical model to compute the capacity \cite{shannon1948mathematical}. 
According to the Shannon-Hartley theorem, the capacity of an additive white Gaussian noise (AWGN) channel can be determined  according to $C=W \log (1+\frac{P}{N_0 W})$ \cite{cover2006elements}, where $W$ is the channel bandwidth, $P$ is the signal power and $N_0$ is the power of AWGN.
Then, the technique of multiple-input and multiple-output (MIMO) was proposed for multiplying the capacity of a wireless channel. For a multi-user (MU)-MIMO channel  with $t$ transmission antennas and $r$ receiving antennas, its uplink channel capacity can be determined according to \cite{telatar1999capacity}, as: \vspace{-0.2em}
\begin{equation}\label{cap_MIMO}
	C=\mbbE \{\log \det (\mbI + \frac{P}{t} \mbH \mbH^*)\}, \vspace{-0.2em}
\end{equation}
where $\mbH$ denotes the channel gain matrix, and $\mbH^*$ is the Hermitian transpose of $\mbH$.
The notation $\mbbE$ represents the expectation.
Note that \eqref{cap_MIMO} is the capacity for a single MU-MIMO channel. Then what is the capacity for a wireless system with multiple MU-MIMO channels? 

For a huge and ultra-dense wireless network, if all the base stations (BSs) are fully cooperated to serve all the users, the signaling overhead will be extremely large and the whole network will become unscalable \cite{Scalable2020, Ubiquitous2019}. To avoid these problems,  the whole network can be divided into multiple clusters \cite{6G2021, yang2021C2, wang2022rcn, deng2022cgn}. Each cluster contains multiple closely-located BSs serving nearby users cooperatively. Then the wireless channel of each cluster can thus be modeled as a MU-MIMO channel. The whole network with multiple non-overlapping clusters can be regarded as a system with multiple MU-MIMO channels, with interference exiting among different channels. According to \cite{yang2021C2}, the average capacity of the $m$-th cluster per BS can be calculated according to
\vspace{-0.3em}
\begin{equation}\label{cap_def_pre}
	C_m \!=\!\mbbE \!\lbrac{ \!\frac{1}{J_m} \!\log \det \left[\mbI \!+\! P(N_0 \mbI \!+\! P \mbPi_m \mbPi_m^*)^{-1}\mbH_m \mbH_m^*\right]\!}\!, 
\end{equation}
where $J_m$ denotes the number of BSs of cluster $m$. $\mbH_m$ is the channel gain matrix of cluster $m$, and $\mbPi_m$ is the interference matrix of cluster $m$, which will be specified later in Section \ref{section2}.  Note that \eqref{cap_def_pre} is more general than \eqref{cap_MIMO}, and thus we will focus on \eqref{cap_def_pre} in the following parts.

There exist many different methods to calculate the capacity based on \eqref{cap_def_pre}.
The most direct way is to calculate the matrix determinant, through conventional schemes, such as singular value decomposition (SVD) and Cholesky decomposition, etc. 
It should be noticed that such conventional determinant-calculation-based methods take a polynomial time to derive the capacity\footnote{More details will be elaborated in  Section \ref{section22}.}. Such a time complexity is obviously  unacceptable for future ultra-dense networks.
Another method, which is applicable for ultra-dense networks,  was developed by Tulino and Verdu \cite{tulino2004random}. They made use of  the random matrix theory (RMT)  to estimate the wireless channel capacity.
However, they mainly focused on  the point-to-point channels with the code-division multiple access (CDMA) scheme,  and their derived capacity expressions are still implicit and complex.
An efficient and general method to derive the capacity  for ultra-dense networks is still missing.

In this paper, we propose a Top-N-Simulated-Estimations (TOSE) algorithm to estimate $C_m$, which is fast, accurate, and general. 
Specifically, we realize the eigenvalue estimation of a large channel gain matrix based on the spiked model in RMT, and then the matrix determinant in \eqref{cap_def_pre} can be  derived with a low complexity. 
As such, the complicated eigenvalue derivation steps (e.g., SVD or Cholesky decomposition) in conventional determinant-calculation-based methods can be totally avoided.
It should be noticed that our TOSE algorithm is designed based on the spike approximations in RMT, which is not utilized in  the conventional RMT-based methods \cite{tulino2004random}. 
Besides of the low complexity,  our algorithm has a high accuracy on capacity estimation, with the estimation error below 5\%. 
Thus, for the ultra-dense wireless networks, where  the number of network nodes (e.g., BSs and users) are large, TOSE has a huge efficiency advantage.
Third, our TOSE algorithm has superior generality, since it is independent of the distribution of network nodes, and the shape of the network area.
\section{System model and Baseline Algorithm}\label{section2}

\subsection{System Model and Capacity Formula}\label{section21}

Consider a wireless network with $J$ single-antenna BSs  and $K$ single-antenna users. The set of BSs is denoted as $\mcS=\{s_1, s_2, \dots s_J \}$, and the set of users is denoted as $\mcU=\{u_1, u_2,\dots,u_K\}$. The network is organized into $M$ non-overlapping clusters, and we use $\mcC_m$ to denote the $m$-th cluster. Then we have $\mcS\bigcup \mcU =\bigcup_{m=1}^M \mcC_m$ \cite{yang2021C2}. 
The sets of the BSs and the users in $\mcC_m$ are denoted by $\mcS_m=\mcS\bigcap \mcC_m$ and  $\mcU_m=\mcU\bigcap \mcC_m$, respectively. 
Moreover, we use $J_m=| \mcS_m|$ and $K_m=| \mcU_m|$ to denote the number of BSs and users in $\mcC_m$. 
In this work, we focus on the ultra-dense scenario, and thus assume $J_m,\;K_m\to\infty\;(m=1,2,\cdots,M)$ \cite{yang2021C2}.

Define the channel gain between the BS $s_j \in \mcS_m$ and the user $u_k \in \mcU$ as 
\begin{equation*}
	\vspace{-0.2em}
	h_{mjk}=l_{mjk} g_{mjk}, \vspace{-0.2em}
\end{equation*}
where $g_{mjk}\sim\mathcal{CN}(0,1)$ is the small-scale fading and 
\begin{equation}\label{l_def}
	l_{mjk}=\left\{\begin{array}{ll}
		d_{mjk}^{-1.75}, & d_{mjk}>d_1, \\
		d_1^{-0.75} d_{mjk}^{-1}, & d_0<d_{mjk}\le d_1, \\
		d_1^{-0.75}d_0^{-1}, & d_{mjk}\le d_0
	\end{array}\right. \vspace{-0.2em}
\end{equation}
is the large-scale fading \cite{wang2022rcn}.
Here, $d_{mjk}$ represents the Euclidean distance between the BS $s_j \in \mcS_m$ and the user $u_k \in \mcU$. 
The parameters $d_0$ and $d_1$ are the near field threshold and far field threshold, respectively.

Thus, we can define the large-scale fading matrix $\mbL_m\in\mbbR^{J_m\times K_m}$ and the small-scaling fading matrix $\mbG_m\in\\ \mbbC^{J_m\times K_m}$, with their $(j,k)$-th entry given by
\begin{equation*}
	[\mbL_m]_{jk}=l_{mjk}, \quad
	[\mbG_m]_{jk}=g_{mjk}, \vspace{-0.2em}
\end{equation*}
where the BS $s_j \in \mathcal{S}_m$ and the user $u_k \in \mathcal{U}_m$. The channel gain matrix $\mbH_m$ in \eqref{cap_def_pre} can thus be defined as \vspace{-0.2em}
\begin{equation}
	\mbH_m=\mbL_m\circ\mbG_m, \vspace{-0.2em}
	\label{eqHP}
\end{equation}
where $\circ$ denotes the Hadamard product. 
The interference matrix $\mbPi_m$ in \eqref{cap_def_pre} can be similarly defined as a Hadamard product of a large-scale fading matrix and a small-scale fading matrix, but these two fading matrices contain the users outside of $\mathcal{C}_m$ and the BSs inside $\mathcal{C}_m$, namely, $u_k \in \mathcal{U}\setminus\mathcal{U}_m$\footnote{$\mathcal{U} \setminus \mathcal{U}_m$ denotes the set of users in $\mathcal{U}$ but not in $\mathcal{U}_m$.}, and $s_j \in \mathcal{S}_m$. Details of deriving $\mbPi_m$ can be found in  \cite{deng2022cgn}, and we omit its elaboration here due to space limitation. To further analyze \eqref{cap_def_pre}, we define
\begin{equation}
	\mbXi_m=N_0 \mbI + P \mbPi_m \mbPi_m^*
	\label{eqXim} \vspace{-0.2em}
\end{equation}
as the noise-plus-interference matrix, with $\mbXi_m\in \mbbC^{J_m \times J_m}$. Based on Lemma 1 in \cite{deng2022cgn}, we know that $\mbXi_m$ converges to a positive definite diagonal matrix as $J_m$ and $K-K_m$ approach infinity, namely, 
\begin{equation}\label{sjj_def}
	\mbXi_m=\textrm{diag}((N_0+P \xi_{11}^m),\dots,(N_0+P \xi_{J_m J_m}^m)),
\end{equation}
where $\xi_{jj}^m=\sum_{ u_k\in \mathcal{U}\setminus\mathcal{U}_m} l_{mjk}^2$. Thus, \eqref{cap_def_pre} can be transformed into \vspace{-0.2em}
\begin{align}
	C_m &=\mbbE\left\{ \frac{1}{J_m}\log \det \sbrac{\mbI + P\mbXi_m^{-1/2} \mbH_m \mbH_m^* \mbXi_m^{-1/2}} \right\} \nonumber \\
	&=\mbbE\left\{ \frac{1}{J_m}\log \det [\mbI + \sbrac{\mbQ_m\circ \mbG_m}\sbrac{\mbQ_m\circ \mbG_m}^*] \right\},\label{cap1} \vspace{-0.2em}
\end{align}
where $\mbQ_m=P^{1/2}\mbXi_m^{-1/2}\mbL_m$. In the following, we will focus on the computation of $C_m$ based on \eqref{cap1}. 

\subsection{Baseline Algorithm Based on Cholesky Decomposition}\label{section22}
It can be observed from \eqref{cap1}  that, the most direct way to calculate $C_m$ is to compute  the logarithm of the determinant of the following matrix \vspace{-0.2em}
\begin{equation*}
	\mbI + (\mbQ_m\circ \mbG_m)(\mbQ_m\circ \mbG_m)^*,
\end{equation*}
and then to obtain its average value. Since the above matrix is Hermitian positive-definite, a classical approach to derive its determinant is to use the Cholesky decomposition, namely, \vspace{-0.2em}
\begin{equation}
	\mbI + (\mbQ_m\circ \mbG_m)(\mbQ_m\circ \mbG_m)^*=\mbR_m\mbR_m^*,
\end{equation}
where $\mbR_m$ is a lower triangular matrix with real and positive diagonal entries $r_{11}^m,\;r_{22}^m,\cdots,r_{J_m J_m}^m$ \cite{press1992numerical}.
As such,  \eqref{cap1} becomes \vspace{-0.2em}
\begin{align}
	C_m \!=\! 2\mbbE\left\{\frac{1}{J_m}\log \det \sbrac{\mbR_m}\right\} =\! 2\mbbE\left\{\frac{1}{J_m}\sum_{j=1}^{J_m}\log r^m_{jj}\right\}. \vspace{-0.2em}
\end{align}

It is well known that the flops of the Cholesky decomposition are $J_m^3/3$ \cite{kress1998numerical}. This is obviously an  unaffordable time complexity for ultra-dense networks, where $J_m$ is extremely large. 
However, such a time complexity is difficult to improve further if we still choose such a  direct method, by calculating the logarithm of the determinant of a large random matrix, to derive the capacity. 
Thus, there is a strong motivation to develop a new method for fast capacity estimation.

\section{TOSE Algorithm Design}\label{section3}
In this section, we will elaborate our TOSE algorithm for fast capacity estimation.
We first propose an approximation of $C_m$, denoted by $\widehat C_m$, by replacing the Hadamard product in \eqref{cap1} with the matrix product.
Second, we obtain an estimation of $\widehat{C}_m$ through fast eigenvalue approximations based on the spiked model in RMT, with which the complicated steps to calculate the exact eigenvalues can be avoided.
Since the top $N$ spiked  eigenvalues can be estimated, we name our algorithm Top-N-Simulated-Estimations (TOSE).

\subsection{Transformation from Hadamard Product to Matrix Product}\label{section31}


It can be observed that there are two Hadamard products in \eqref{cap1}. Unfortunately, using RMT to analyze the Hadamard product of large-dimensional random matrices is difficult and lack of closed-form expressions \cite{silverstein2021}.  
Thus, our first step is to transform the Hadamard product to the classic  matrix product. 
We propose a method to optimally replace the Hadamard product $\mbQ_m \circ \mbG_m$ by the matrix product $\mbT_m \mbG_m$,  so that we can obtain an approximated expression of $C_m$ as
\begin{equation}\label{cap2}
	\widehat{C}_m=\frac{1}{J_m}\mbbE\Big\{\log \det \big(\mbI +\mbT_m\mbG_m\mbG_m^{*}\mbT_m^*\big)\Big\}. \vspace{-0.2em}
\end{equation}
Here, the matrix $\mbT_m$ is diagonal, and its $j$-th diagonal entry equals to the average value over all the entries at the $j$-th row  of $\mbQ_m$, namely, 
\begin{equation}\label{T_def}
	\mbT_m=\mathrm{diag}(t_{m1}, \dots, t_{m J_m}),  \vspace{-0.2em}
\end{equation}
and \vspace{-0.2em}
\begin{equation*}
	t_{mj}=\frac{1}{K_m}\sum_{k=1}^{K_m} q_{mjk},  
\end{equation*}
and $q_{mjk}$ is the $(j,k)$-th entry of $\mbQ_m$, i.e., $ [\mbQ_m]_{jk}=q_{mjk}$.
Such a transformation  provides the basis of our TOSE algorithm design, and its optimality  is proved in the following theorem.
\medskip
\begin{theorem}\label{theorem1}
	For any matrix $\widetilde{\mbT}_m$, define 
	\begin{equation*}
		\mbDel_m=\mbQ_m\circ \mbG_m - \widetilde{\mbT}_m\mbG_m \vspace{-0.2em}
	\end{equation*}
	and 
	\begin{equation*}
		\quad F_m=\mbbE \big(\norm{\mbDel_m}_F^2\big). \vspace{-0.2em}
	\end{equation*}
	$F_m$ is minimized if and only if $\widetilde{\mbT}_m$ is \eqref{T_def}, and the minimum $F_m$ is
	\begin{equation}
		F_m|_{\min}= \sum_{j=1}^{J_m} \left[\sum_{k=1}^{K_m} q_{mjk}^2 - \frac{1}{K_m}\left(\sum_{k=1}^{K_m} q_{mjk}\right)^2\right]. \vspace{-0.2em}
	\end{equation}
\end{theorem}
\smallskip
\begin{IEEEproof}
	Consider the second absolute moment of \\ $[\mbDel_m]_{jk}=\delta_{mjk}$, it writes \vspace{-0.2em}
	\begin{align}
		\mbbE\lbrac{\abs{\delta_{mjk}}^2}
		&=\mbbE\lbrac{\abs{q_{mjk}g_{mjk}-\sum_{n=1}^{J_m} \tilde{t}_{mjn}g_{mnk}}^2} \nonumber\\
		&=q_{mjk}^2 + \sum_{n=1}^{J_m} \tilde{t}_{mjn}^2 - 2\tilde{t}_{mjj}q_{mjk}. \label{Delta_jk} \vspace{-0.2em}
	\end{align}
	Thus, we have \vspace{-0.2em}
	\begin{equation}\label{def_F}
		\begin{split}
			F_m=&\mbbE \big(\norm{\mbDel_m}_F^2\big)
			= \sum_{j,k}\mbbE \big(\abs{\delta_{mjk}}^2\big) \\
			=&\sum_{j=1}^{J_m}\sum_{k=1}^{K_m} q_{mjk}^2 
			\!+\! K_m \sum_{j=1}^{J_m}\sum_{n=1}^{J_m} \tilde{t}_{mjn}^2
			\!-\! 2\sum_{j=1}^{J_m}\sum_{k=1}^{K_m} \tilde{t}_{mjj}q_{mjk}. \vspace{-0.2em}
		\end{split}
	\end{equation}
	The above formula is the quadratic function of each $\tilde{t}_{mjn}$, so $F_m$ reaches its minimum value when \vspace{-0.2em}
	\begin{equation*}
		\tilde{t}_{mjn}=\begin{cases}
			0, &j\ne n; \\ \displaystyle\frac{1}{K_m}\sum_{k=1}^{K_m} q_{mjk}, & j=n.
		\end{cases} \vspace{-0.2em}
	\end{equation*}
	This is exactly the form given in \eqref{T_def}. By substituting  the above formula into \eqref{def_F}, we obtain \vspace{-0.2em}
	\begin{equation}
		F_m|_{\min}= \sum_{j=1}^{J_m} \left[\sum_{k=1}^{K_m} q_{mjk}^2 - \frac{1}{K_m}\left(\sum_{k=1}^{K_m} q_{mjk}\right)^2\right]. \vspace{-0.3em}
	\end{equation}
\end{IEEEproof}

\subsection{TOSE Algorithm  Based on Spike Approximations}\label{section32}
Based on  the analysis of  Section \ref{section31}, we now focus on \eqref{cap2}, to design a fast algorithm to estimate $\widehat{C}_m$. 

Define \vspace{-0.2em}
\begin{equation} \label{Bm}
	\mbB_m = \mbT_m \mbG_m \mbG_m^* \mbT_m^*\in \mbbC^{J_m\times J_m}, \vspace{-0.2em}
\end{equation}
and  thus (\ref{cap2}) can be written as
\begin{equation}
	\widehat{C}_m=\frac{1}{J_m}\mbbE \left\{\log \det \big(\mbI + \mbB_m\big)\right\}.
	\label{eqAC1} \vspace{-0.2em}
\end{equation}
Note that the rank of $\mbB_m$ follows
\begin{equation*}
	\textrm{rank}(\mbB_m)\le \min(J_m,K_m),
\end{equation*}
because of $\mbG_m \in \mbbC^{J_m\times K_m}$.
Moreover, the  eigenvalue decomposition of $\mbI+\mbB_m$ can be  given by
\begin{align}
	\mbI+\mbB_m=\mbU_m\mbSg_m\mbU_m^*,
\end{align}
where $\mbU_m$ is a $J_m\times J_m$ complex unitary matrix, and
\begin{equation*}
	\mbSg_m=\textrm{diag}(\sigma_1,\sigma_2,\cdots,\sigma_{J_m}). \vspace{-0.2em}
\end{equation*}
Here, $\sigma_1\ge\cdots\ge\sigma_{J_m}\ge 0$ are the eigenvalues of $\mbSg_m$, and $\widehat C_m$ in \eqref{eqAC1} is equivalent to \vspace{-0.2em}
\begin{equation} \label{cap3}
	\widehat{C}_m = \frac{1}{J_m} \mbbE\left\{ \sum\limits_{j=1}^{J_m} \log \sbrac{{\sigma}_j}\right\}. \vspace{-0.2em}
\end{equation}

As analyzed in Section \ref{section22}, the existing method to calculate the eigenvalues (e.g., $\sigma_j$) takes a polynomial time complexity $O(J_m^3)$.  In the following, we will use the spiked model in RMT to realize the fast eigenvalue estimations. Based on  RMT, the eigenvalue calculations of a large-dimensional random matrix can be simplified by its limiting spectral distribution \cite{MP1967}. To approximate $\sigma_j$ by the spiked model in RMT, we first randomize the matrix $\mbI+\mbB_m$ by replacing the population identity matrix $\mbI$ with its corresponding sample covariance matrix $\frac{1}{K_m}\widetilde{\mbG}_m\widetilde{\mbG}_m^*$,
which  is a commonly used approach in statistical inference. The random matrix $\widetilde{\mbG}_m\in \mbbC^{J_m\times K_m}$ has independent and identically distributed (i.i.d.) entries with zero mean and unit variance. As such,
$\widehat C_m$ in \eqref{eqAC1} can be approximated by
\begin{equation} \label{cap4}
	\widehat{C}_m \approx \frac{1}{J_m} \mbbE \left\{\log\det \left(\frac{1}{K_m}\widetilde{\mbG}_m\widetilde{\mbG}_m^* + \mbB_m \right)\right\} \vspace{-0.2em}
\end{equation}
Note that the following matrix 
\begin{equation}
	\frac{1}{K_m}\widetilde{\mbG}_m\widetilde{\mbG}_m^* + \mbB_m
	\label{GtildeB} \vspace{-0.2em}
\end{equation}
can be regarded as a large-dimensional spiked random matrix. The properties of  \eqref{GtildeB} is thus  dominated by its largest $N$ eigenvalues,  which can be called the spikes. 
Then we can use the limiting spectral distribution of large-dimensional spiked matrix \cite{bai2010spectral} to get the approximated eigenvalues of the matrix \eqref{GtildeB}.


If we denote all the $J_m$ approximated eigenvalues of matrix (\ref{GtildeB}) as
\begin{equation*}
	\tilde{\sigma}_1\ge \tilde{\sigma}_2\ge \cdots\ge \tilde{\sigma}_N\ge \tilde{\sigma}_{N+1}\ge \cdots\ge \tilde{\sigma}_{J_m}, \vspace{-0.2em}
\end{equation*}
with $N$ much larger eigenvalues, 
$\widehat{C}_m$ in \eqref{cap4} can be estimated by \vspace{-0.2em}
\begin{equation} \label{cap_final}
	\widehat{C}_m \!
	\approx\!  \frac{1}{J_m}\mbbE \left\{ \sum\limits_{j=1}^{N} \log \sbrac{\tilde{\sigma}_j} \right\}.	
\end{equation}

Next, we will show how to efficiently get the approximated top $N$ eigenvalues, which are \vspace{-0.2em}
\begin{equation*}
	\tilde{\sigma}_1, \; \tilde{\sigma}_2, \cdots, \tilde{\sigma}_N. \vspace{-0.2em}
\end{equation*}
Based on RMT,  $\tilde{\sigma}_1, \cdots, \tilde{\sigma}_N$ can be treated as the approximations of the $N$ largest  spiked eigenvalues, which locate outside of the supporting set of the standard  Mar\v{c}enko-Pastur (MP) law \cite{Johnstone2001}. 
According to the standard MP-law, we know that these eigenvalues are bounded and can be denoted by \vspace{-0.2em}
\begin{equation*}
	\theta_2>\tilde{\sigma}_1\ge \tilde{\sigma}_2\ge \cdots\ge \tilde{\sigma}_N>\theta_1, \vspace{-0.2em}
\end{equation*}
as $J_m,\;K_m\to\infty$ and $K_m/J_m \to \beta$. 
The lower bound $\theta_1$ can be determined by \vspace{-0.3em}
\begin{equation} \label{lowbound}
	\theta_1=\sbrac{1+1/\sqrt{\beta}}^2. \vspace{-0.2em}
\end{equation}
The exact value of the upper bound $\theta_2$  will not be utilized in our algorithm and thus we can omit its derivation here.
By assuming that $\tilde{\sigma}_1, \; \tilde{\sigma}_2, \cdots, \tilde{\sigma}_N$ are evenly spaced over the interval $[\theta_1,\;\theta_2]$ with space $\Delta \sigma$, we have
\begin{equation} \label{sigma_j}
	\tilde{\sigma}_j\!=\! \theta_1+(N\!+\!1\!-\!j)\Delta \sigma, \quad j=1,2,\cdots,N,
\end{equation}
and
\begin{equation*}
	\sum_{j=1}^N \tilde{\sigma}_j -N=\textrm{tr}(\mbB_m), \vspace{-0.2em}
\end{equation*}
where $\textrm{tr} (\mbB_m)$ denotes the trace of matrix $\mbB_m$. Based on the above formulas, we can obtain 
\begin{equation}\label{step_len}
	\Delta \sigma=\frac{2[\textrm{tr} (\mbB_m) +N-N\theta_1]}{N(N+1)}.
\end{equation}

By substituting \eqref{lowbound} and \eqref{step_len} into \eqref{sigma_j}, we can directly approximate the top $N$ eigenvalues of \eqref{GtildeB}, namely, $\tilde{\sigma}_j (j = 1,2,..., N)$.
By substituting $\tilde{\sigma}_j$ into \eqref{cap_final}, the approximated capacity $\widehat {C}_m$ can be obtained.
The above procedures are summarized in the following Algorithm \ref{tose_algorithm}, which is TOSE.
\begin{figure}[htbp] 
	\vspace{-2.5em}
	\renewcommand{\algorithmicrequire}{\textbf{Input:}}
	\renewcommand{\algorithmicensure}{\textbf{Output:}}
	\removelatexerror
	\begin{algorithm}[H]
		\caption{Top $N$ simulated estimations (TOSE)}
		\begin{algorithmic}[1]
			\REQUIRE $\mbB_m$, $N$
			\ENSURE Estimation of $\widehat C_m$.
			\STATE Calculate $\mathrm{tr} (\mbB_m)$.
			\STATE Calculate  $\Delta \sigma=\frac{2[\mathrm{tr} (\mbB_m) +N-N\theta_1]}{N(N+1)}$, where $N$ is the number of the spikes and $\theta_1=(1+1/\sqrt{\beta})^2$.
			\STATE Compute 
			$\tilde\sigma_j=\theta_1+(N+1-j)\Delta \sigma,\; j=1,\cdots, N$.
			\STATE Compute $ \displaystyle\frac{1}{J_m} \sum\limits_{j=1}^{N} \log (\tilde\sigma_j) $ as an estimation of $\widehat C_m$.
			
		\end{algorithmic}
		\label{tose_algorithm}
	\end{algorithm}
	\vspace{-1.5em}
\end{figure}

Note that in  Algorithm \ref{tose_algorithm},  the matrix $\mbB_m$ is given, the complexity
for the TOSE itself is only $O(J_m)$. Therefore, the complexity of TOSE can be   linear.
If $\mbB_m$ is not given,  $ \mathrm{tr} (\mbB_m) $ can be determined directly according to 
\begin{equation}\label{trB}
	\mathrm{tr} (\mbB_m) = \sum_{j=1}^{J_m} b_{mjj} = \sum_{j=1}^{J_m}\sum_{k=1}^{K_m} t_{mjk}^2 |g_{mjk}|^2,
\end{equation}
whose time complexity is $O(J_m^2)$, which is still less than $O(J_m^3)$.

\section{Performance Evaluation}\label{section4}

In this section, we perform simulations to illustrate the high accuracy, the superior generality, and the low complexity of our TOSE algorithm on capacity estimations.

\begin{figure}[ht]
	\centering
	\vspace{-1em}
	\subfloat[]{\includegraphics[width=0.5\linewidth, keepaspectratio]{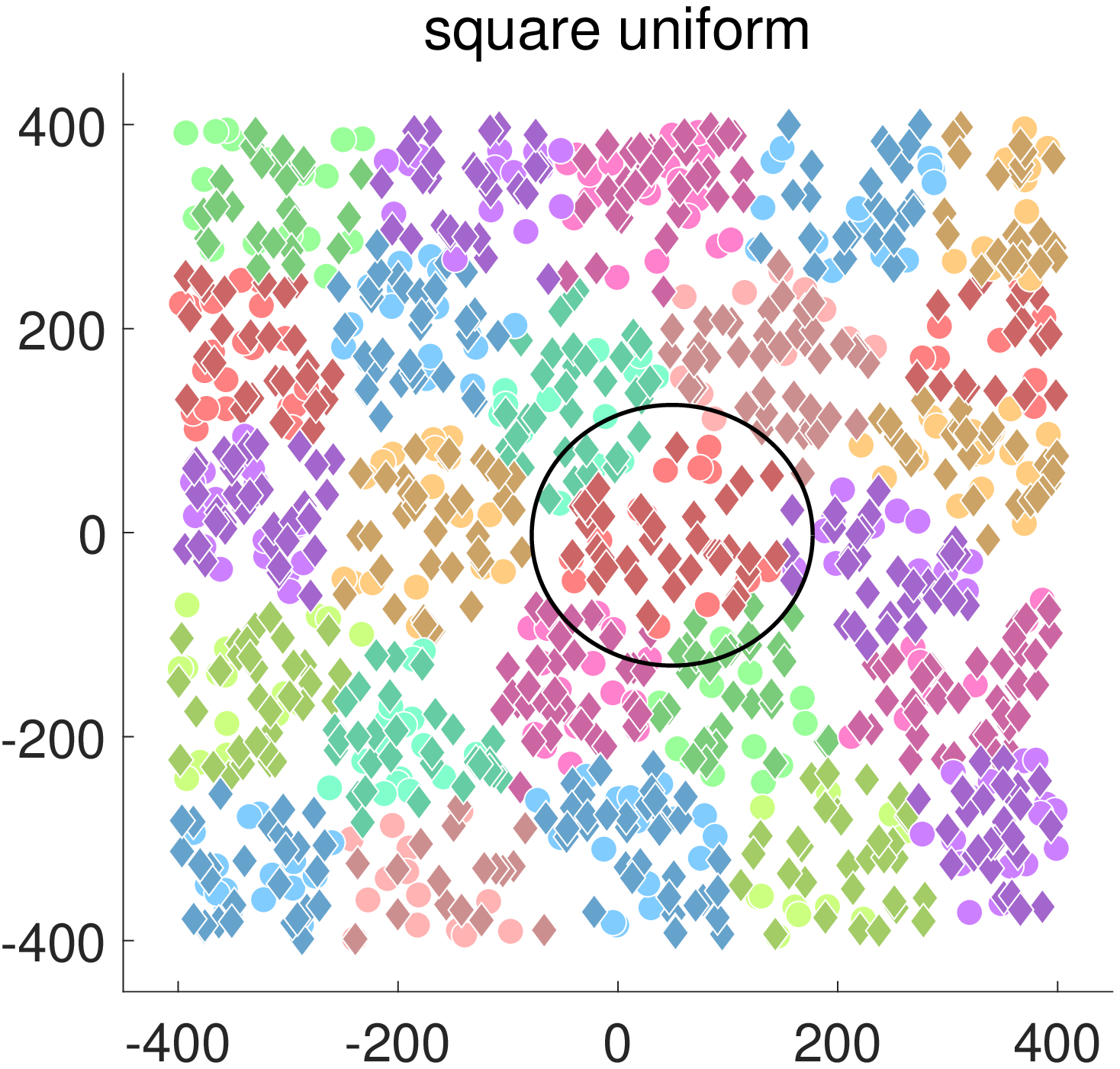}}
	\hspace{-0.7em}
	\subfloat[]{\includegraphics[width=0.5\linewidth, keepaspectratio]{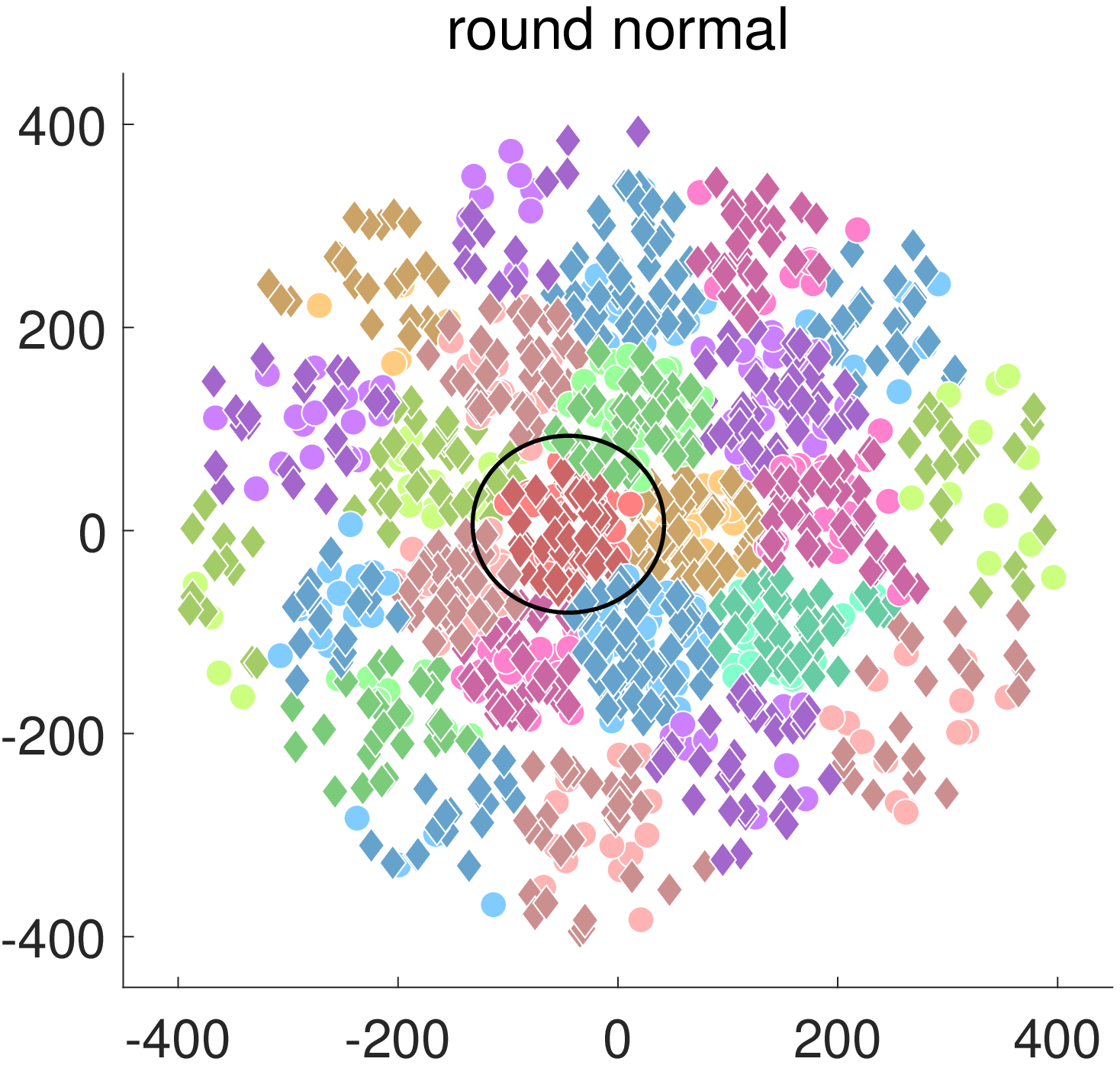}}
	\vspace{-0.5em}
	\caption{Illustration of two different scenarios of  ultra-dense networks. Each color represents an individual cluster. The brighter dots represent the BSs and the darker diamonds represent the users. We randomly pick one cluster (marked by a black circle) as our focus to analyze the capacity. (a) Square network area with uniformly
		distributed network nodes. (b) Round network area with truncated normally distributed network nodes.}
	\vspace{-1em}
	\label{net_plot}
\end{figure}

We consider the following two different ultra-dense wireless scenarios to reflect the generality of TOSE, which are:   

\begin{itemize}
	\item[(a)] A square network area (with side length $D$), and the uniformly distributed network nodes (BSs and users);
	\item[(b)] A round network area (with diameter $D$), and the truncated normally distributed network nodes.
\end{itemize}
The corresponding schematic diagrams  are given in Fig. \ref{net_plot}, with parameter settings listed in Table \ref{parameters}. To reduce the influence caused by the randomness, we generate $200$ random experiments for each scenario, and compute the average capacity of a randomly picked cluster (marked by a black circle in Fig. \ref{net_plot}) as an example case. 
Before that, we use the k-means algorithm to simulate the networks with non-overlapping clusters \cite{deng2022cgn}.
All the experiments are conducted on a platform with 16G RAM, and a Intel(R) Core(TM) i5-10400 CPU @2.90GHz with 6 cores.
The program was locked on a single thread to avoid the influence of multi-thread acceleration.

\begin{figure*}[bp]
	\centering
	\vspace{-1em}
	\subfloat[]{\includegraphics[width=0.25\linewidth]{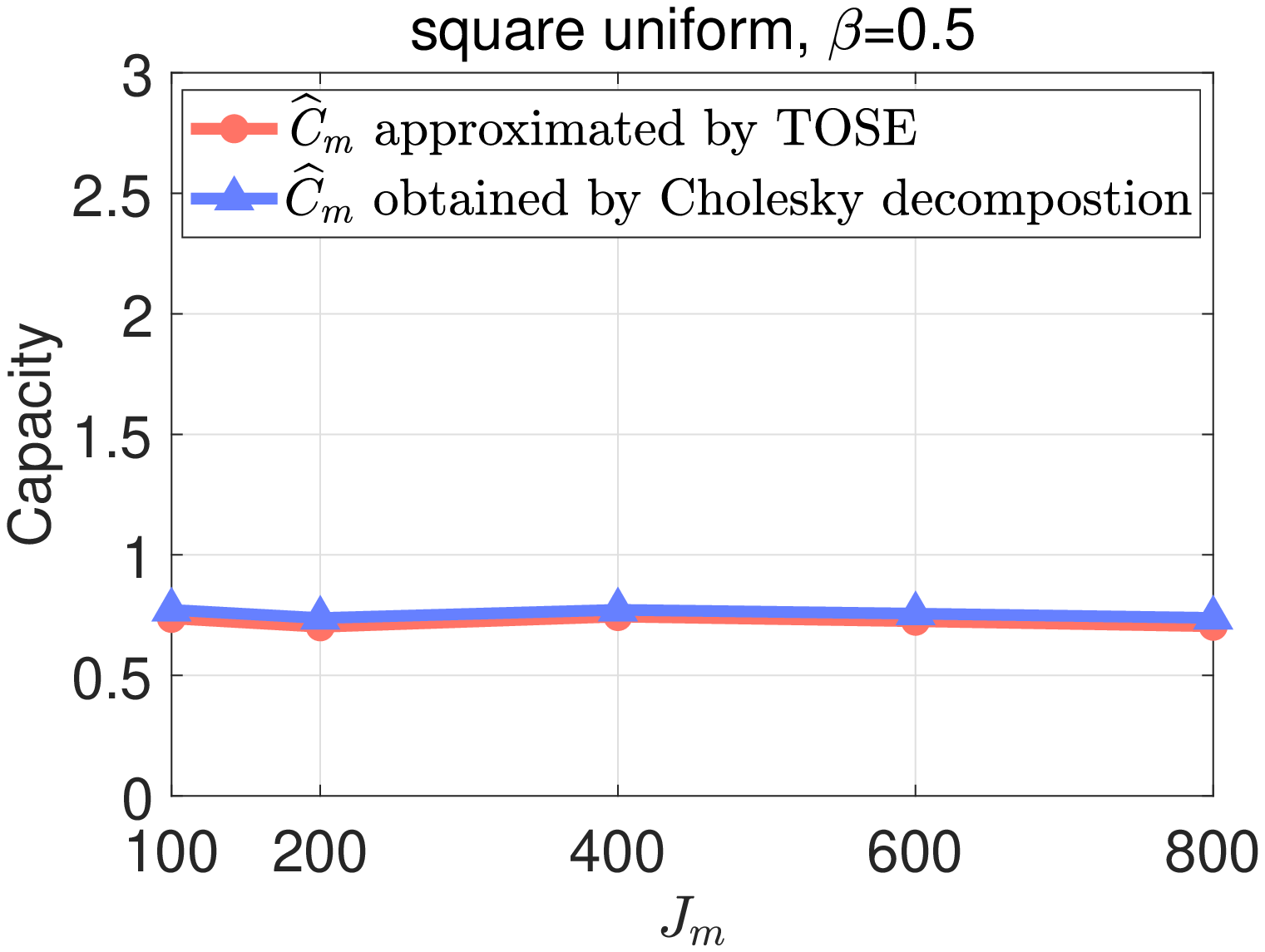}} \hspace{-0.7em}
	\subfloat[] {\includegraphics[width=0.25\linewidth]{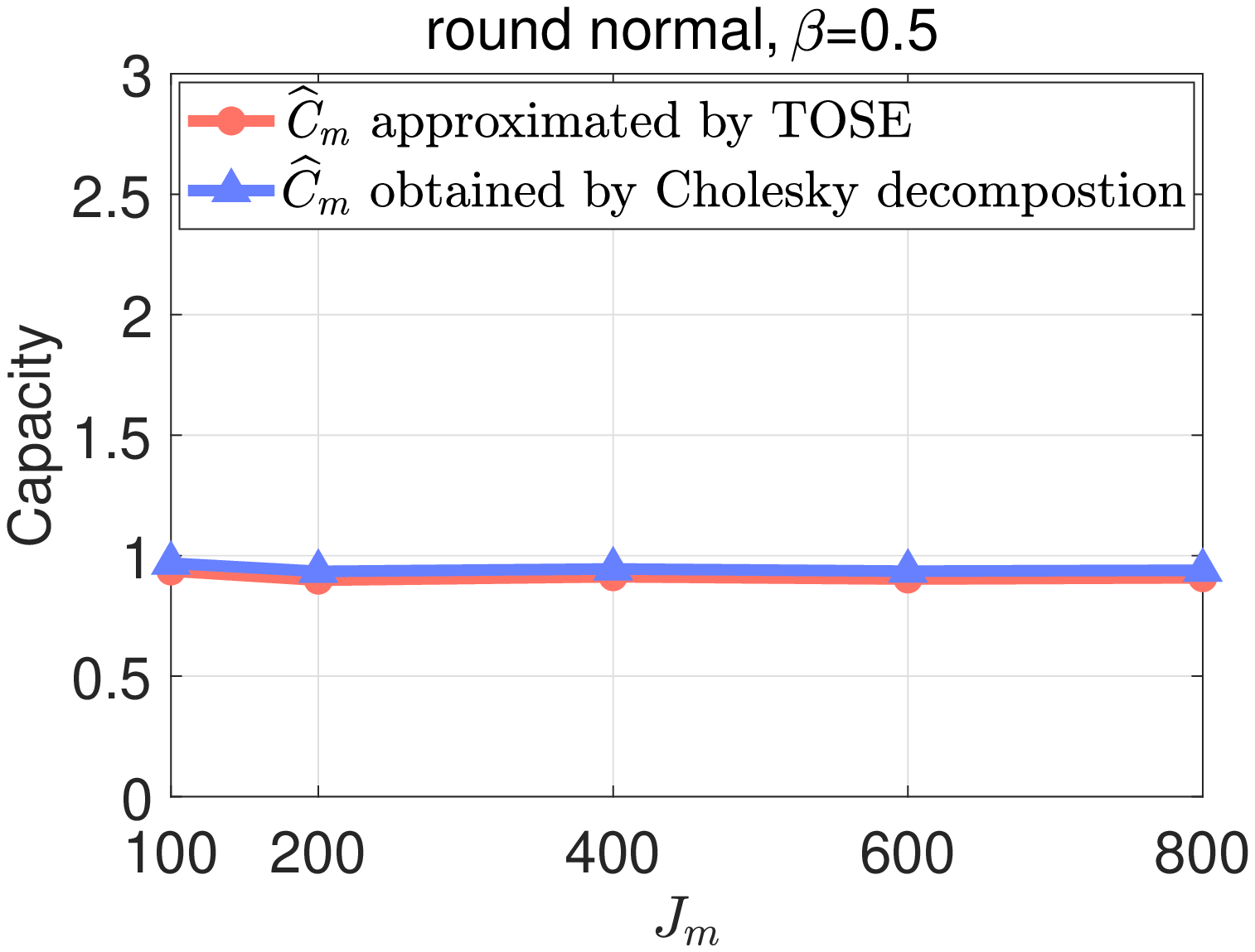}} \hspace{-0.7em}
	\subfloat[]{\includegraphics[width=0.25\linewidth]{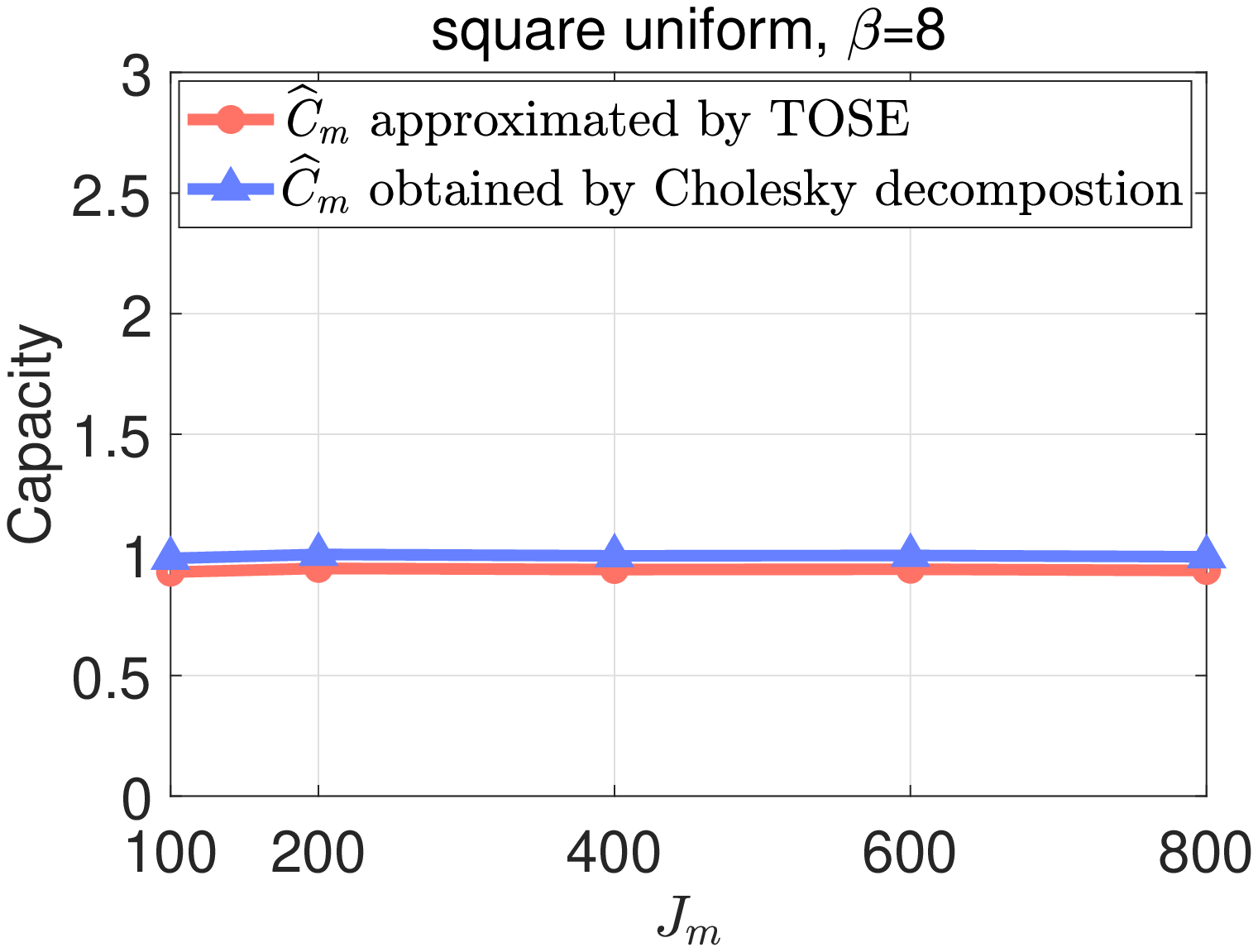}} \hspace{-0.7em}
	\subfloat[] {\includegraphics[width=0.25\linewidth]{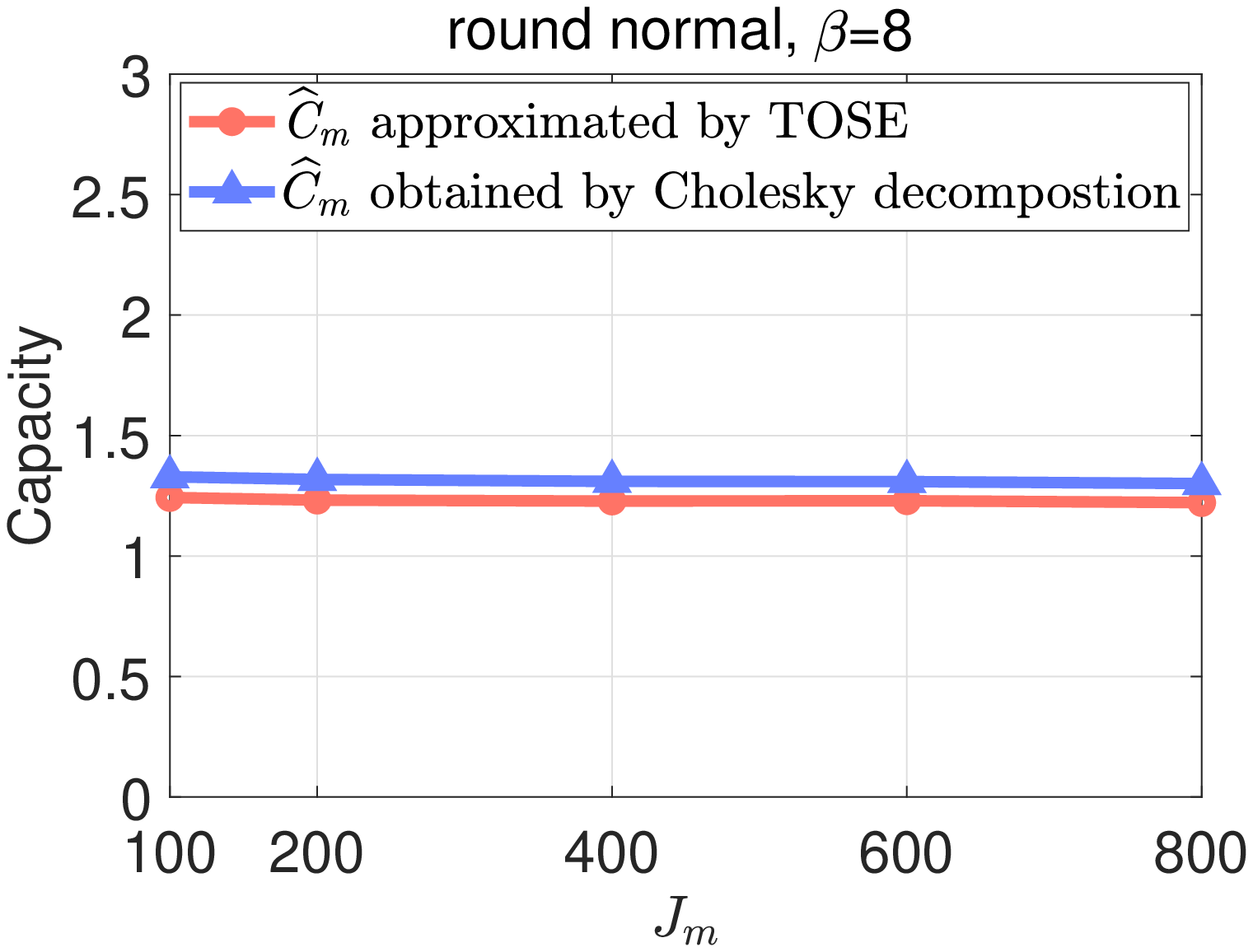}}
	\vspace{-0.5em}
	\caption{The comparison of $\widehat C_m$ obtained by the TOSE algorithm (red lines) and the benchmark algorithm based on Cholesky decomposition (blue lines) under different network settings.
		(a) and (c): Network nodes are uniformly distributed in a square network area,  with $\beta=0.5$ and $\beta=8$, respectively. (b) and (d): Network nodes are truncated normally distributed in a round network area, with $\beta=0.5$  and $\beta=8$, respectively.}
	\label{tose_results}
\end{figure*}
\begin{figure*}[ht]
	\centering
	\subfloat[]{\includegraphics[width=0.25\linewidth,keepaspectratio]{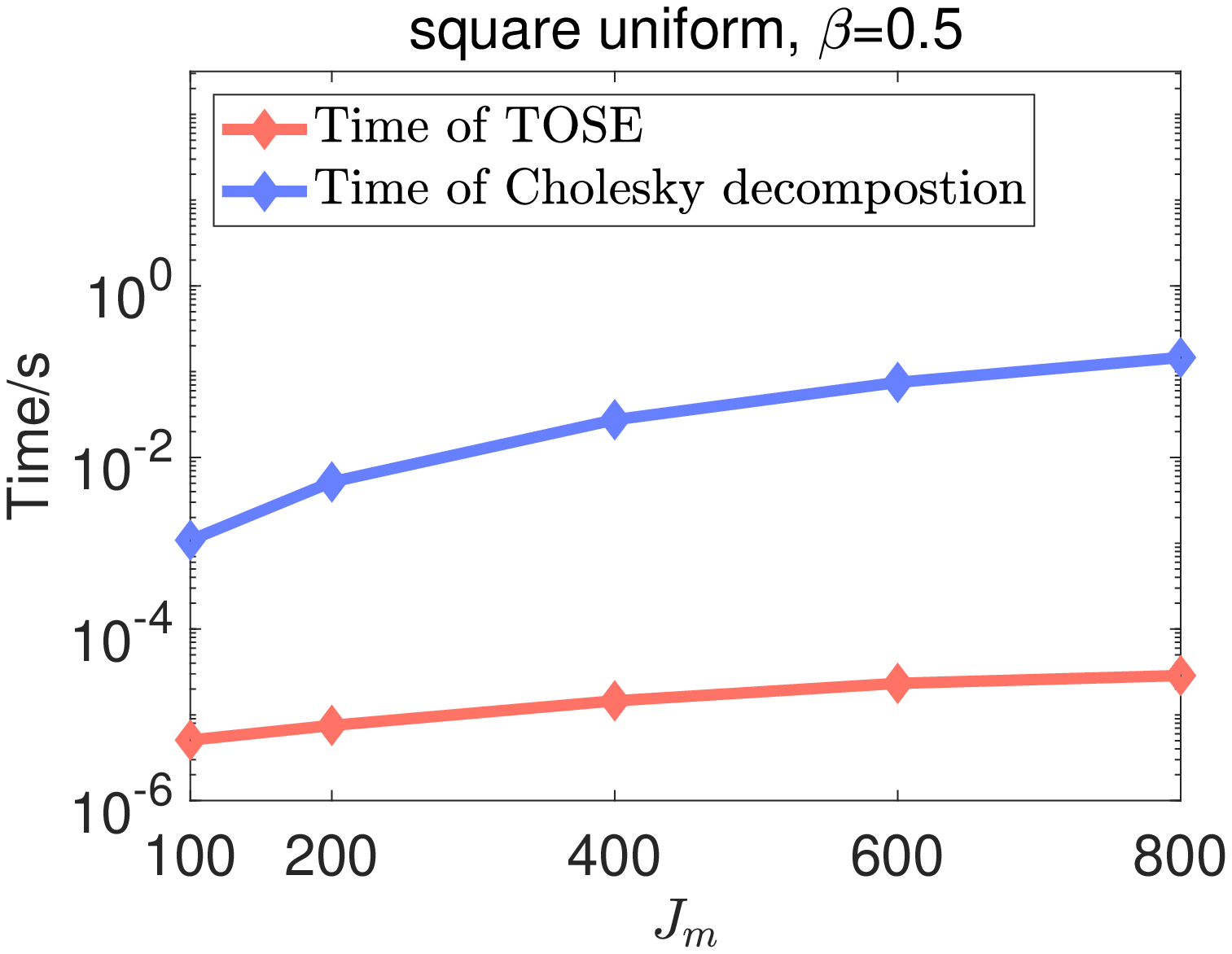}} \hspace{-0.7em}
	\subfloat[] {\includegraphics[width=0.25\linewidth,keepaspectratio]{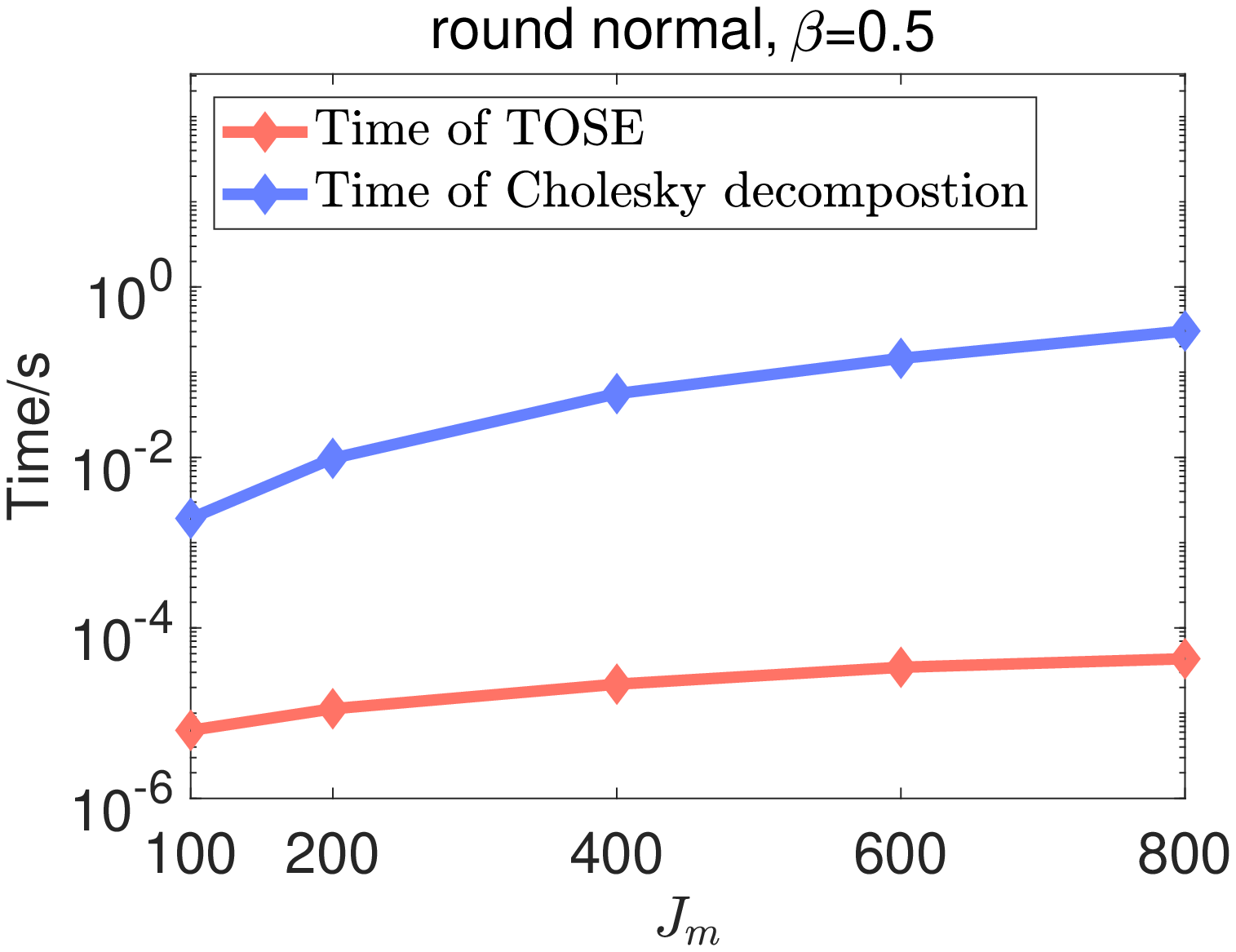}} \hspace{-0.7em}
	\subfloat[]{\includegraphics[width=0.25\linewidth,keepaspectratio]{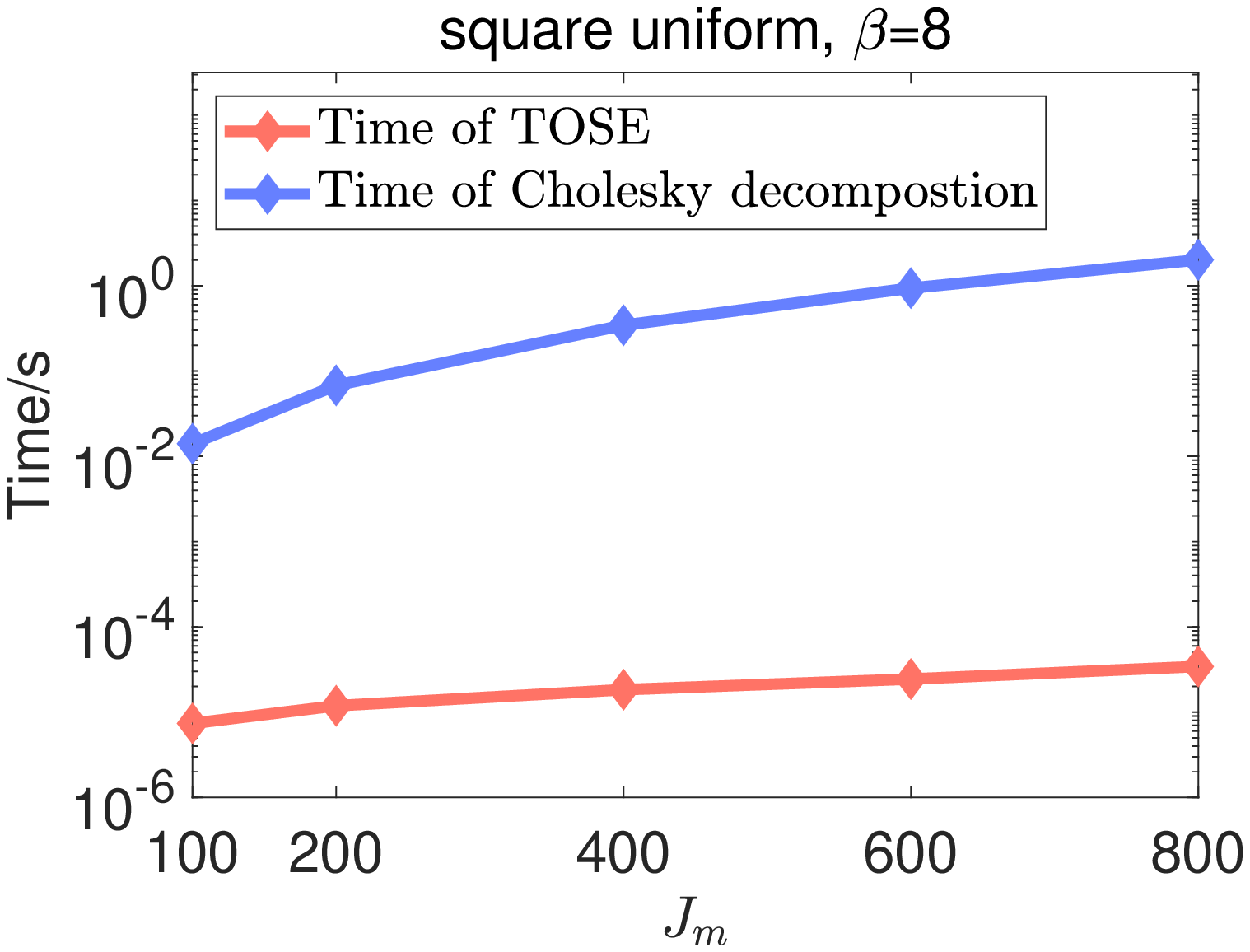}} \hspace{-0.7em}
	\subfloat[] {\includegraphics[width=0.25\linewidth,keepaspectratio]{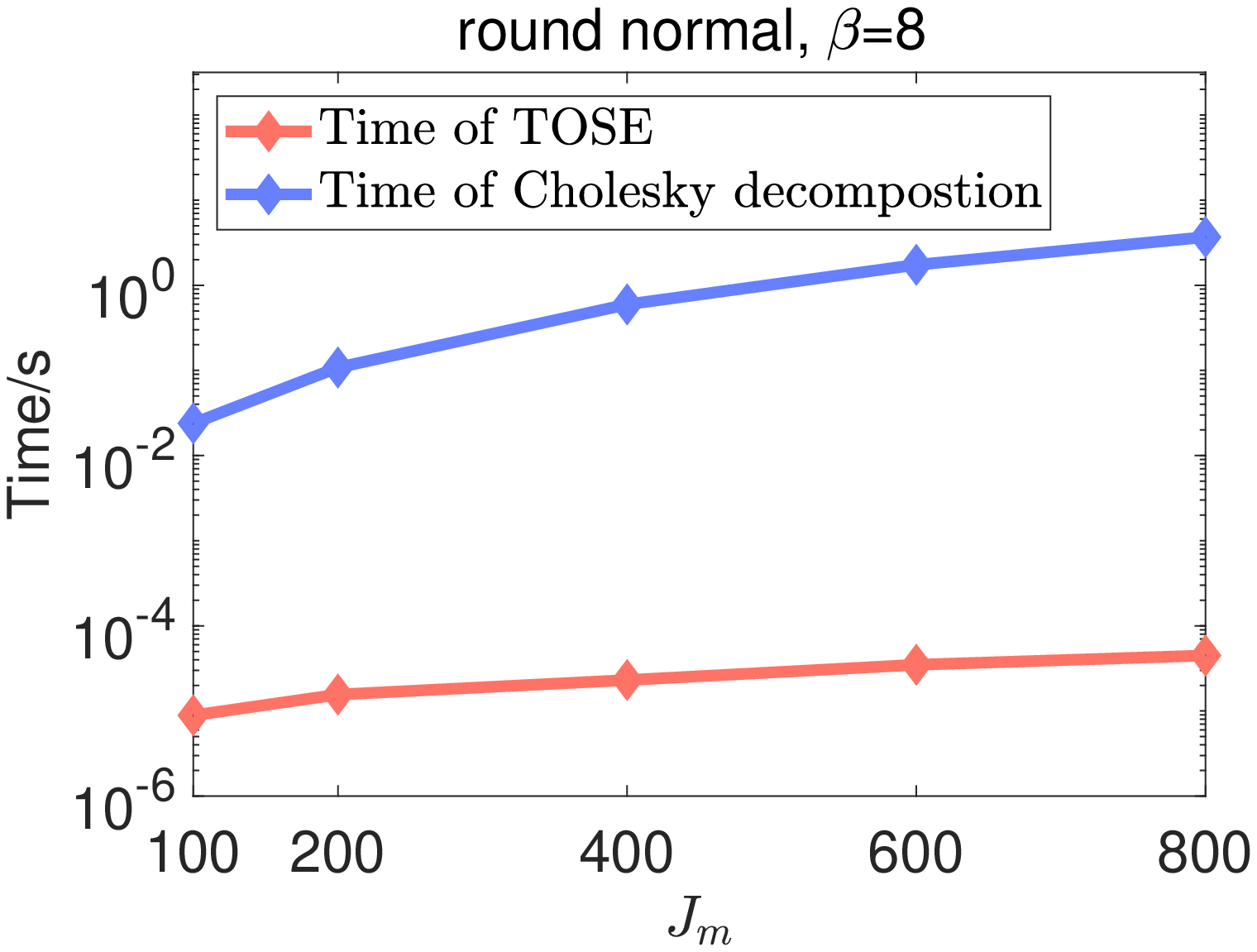}}
	\vspace{-0.5em}
	\caption{The comparison of computational time by the TOSE algorithm (red lines) and the benchmark algorithm based on Cholesky decomposition (blue lines) under different network settings.
		(a) and (c): Network nodes are uniformly distributed in a square network area,  with $\beta=0.5$  and $\beta=8$, respectively. (b) and (d): Network nodes are truncated normally distributed in a round network area, with $\beta=0.5$ and $\beta=8$, respectively.}
	\label{tose_time}
\end{figure*}

\begin{figure*}[ht]
	\centering
	\subfloat[]{\includegraphics[width=0.25\linewidth,keepaspectratio]{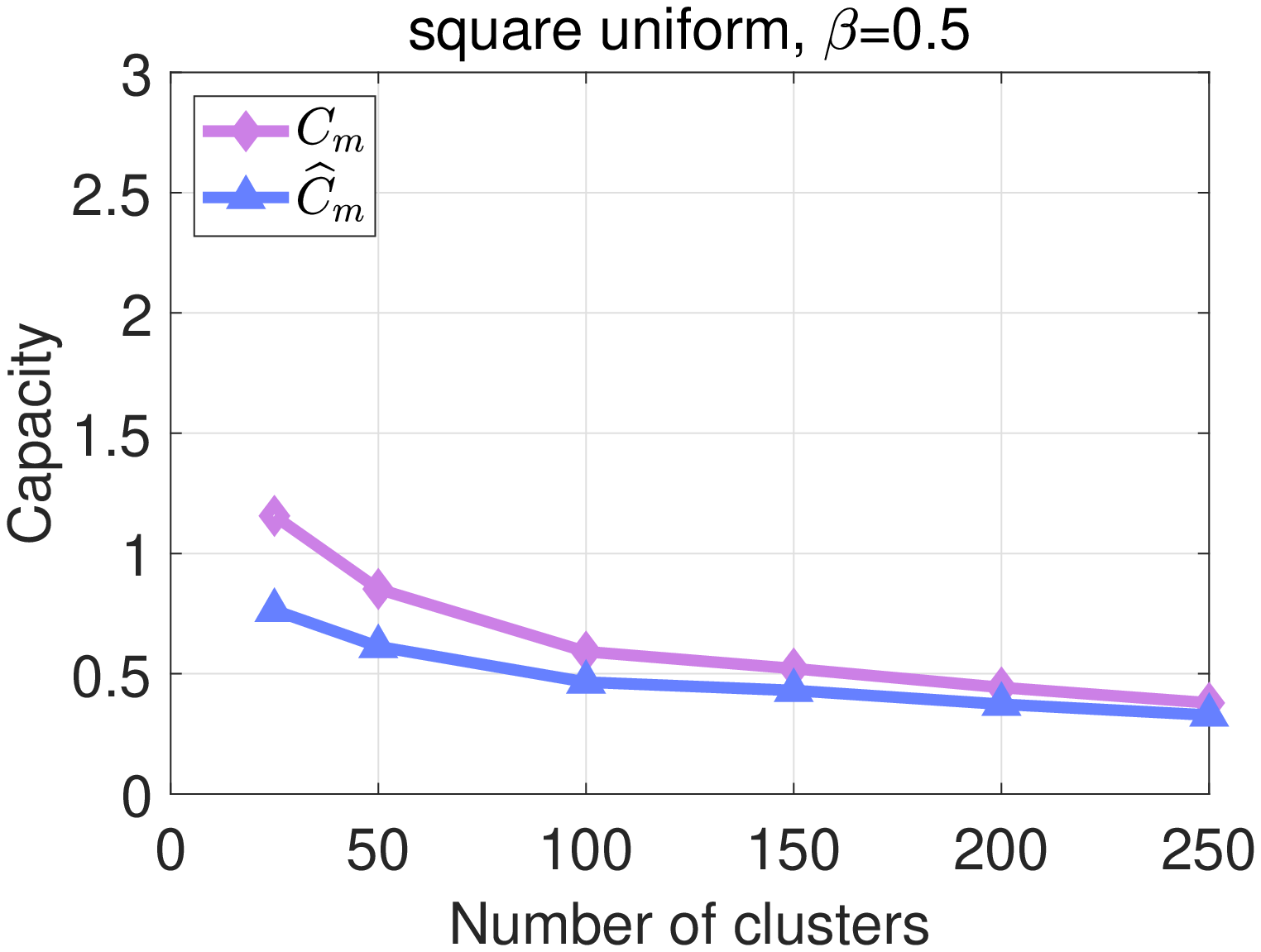}} \hspace{-0.7em}
	\subfloat[] {\includegraphics[width=0.25\linewidth,keepaspectratio]{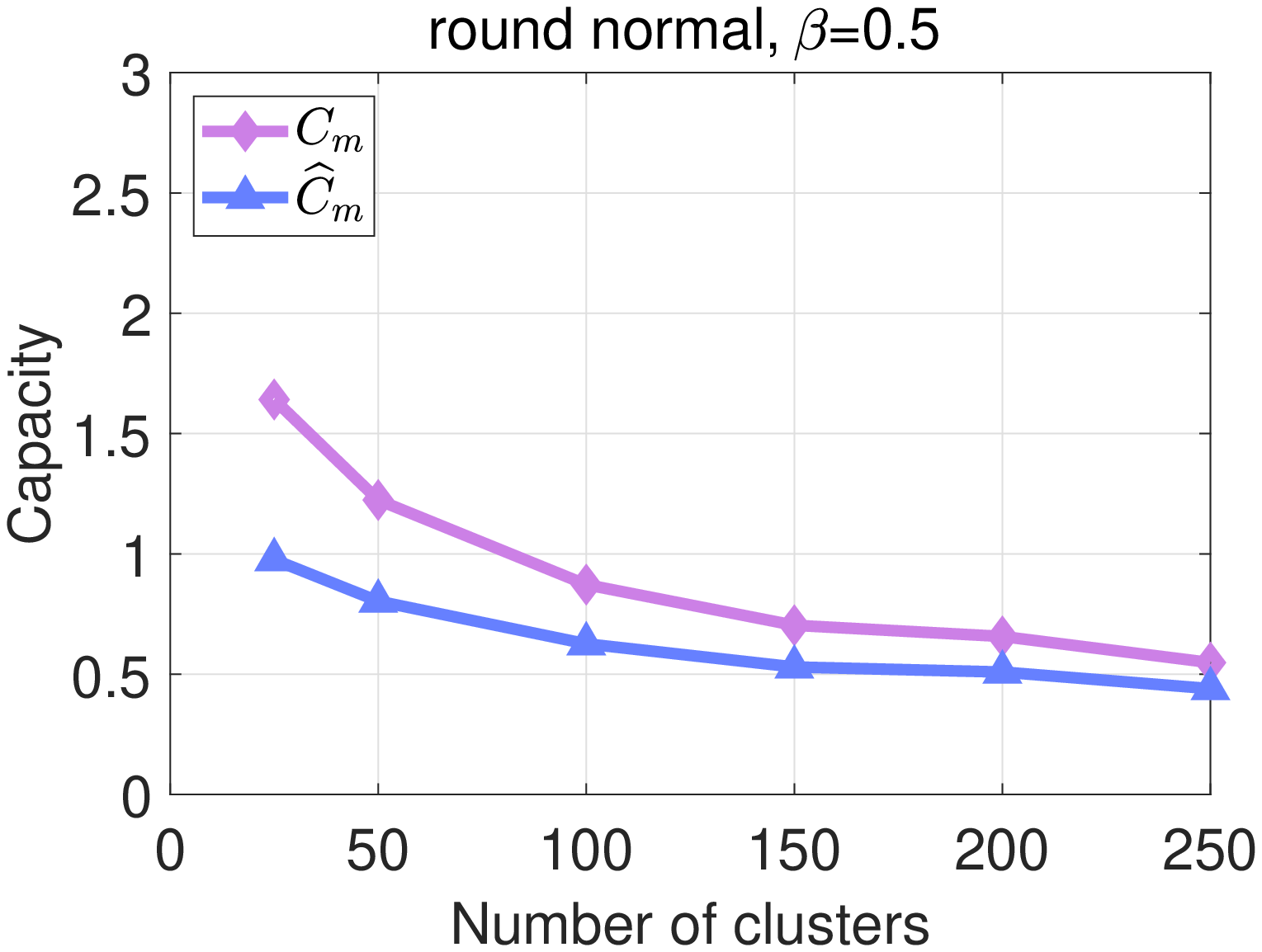}} \hspace{-0.7em}
	\subfloat[]{\includegraphics[width=0.25\linewidth,keepaspectratio]{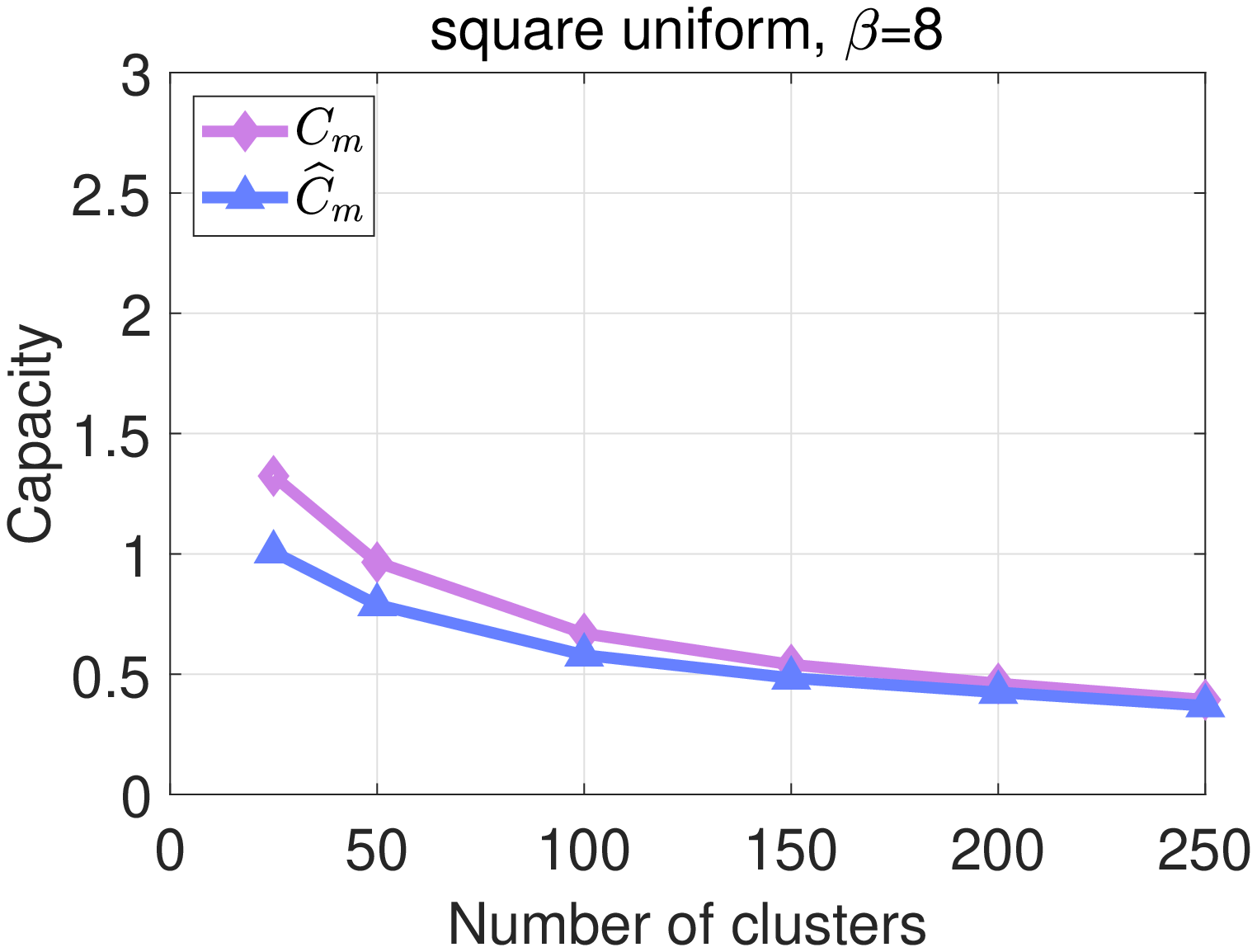}} \hspace{-0.7em}
	\subfloat[] {\includegraphics[width=0.25\linewidth,keepaspectratio]{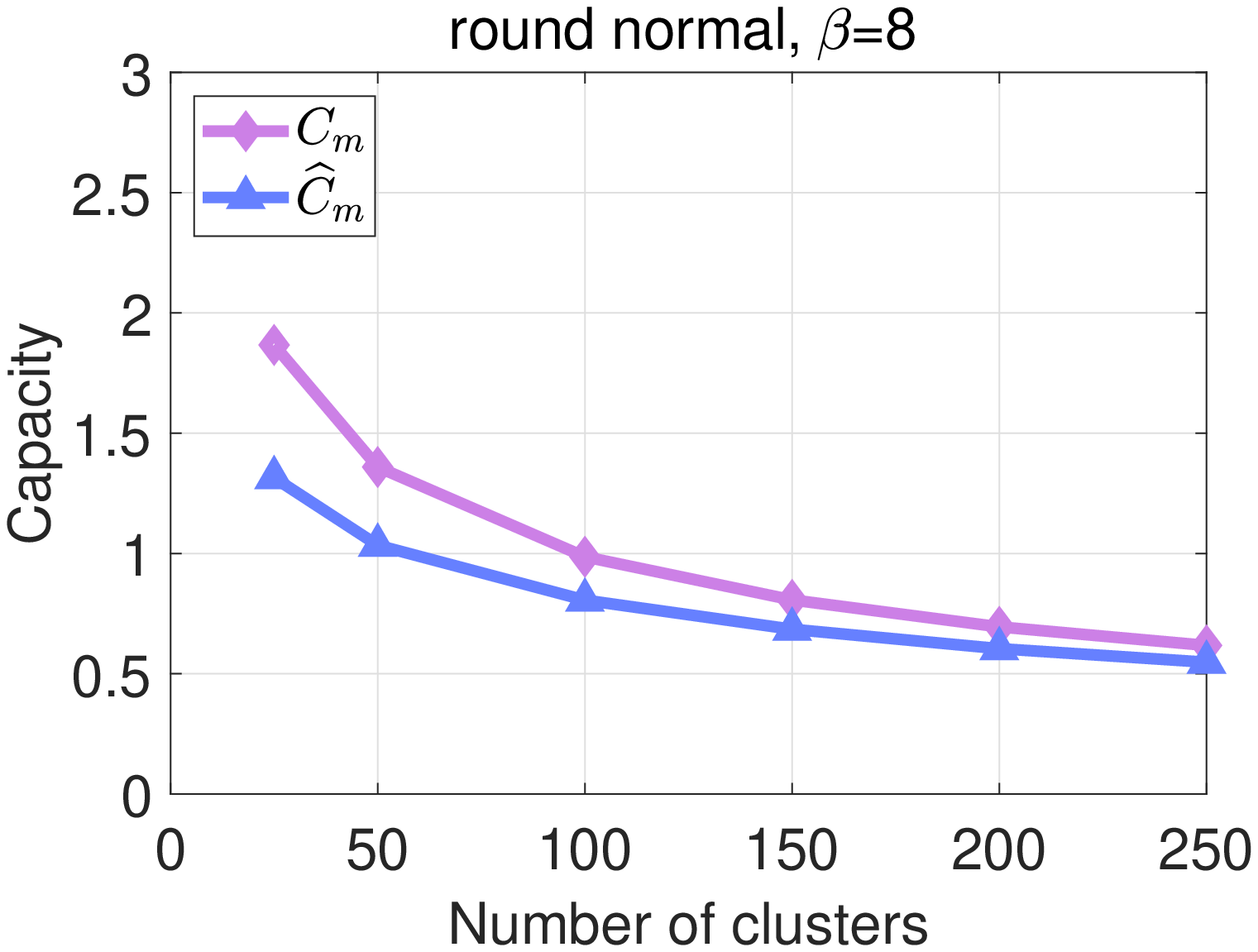}}
	\vspace{-0.5em}
	\caption{Comparisons between  the theoretical $ C_m$ (purple lines)  and approximated $\widehat C_m$ (blue lines) under different network settings.
		(a) and (c): Network nodes are uniformly distributed in a square network area,  with $\beta=0.5$  and $\beta=8$, respectively. (b) and (d): Network nodes are truncated normally distributed in a round network area, with $\beta=0.5$  and $\beta=8$, respectively. }
	\vspace{-1em}
	\label{real_appro}
\end{figure*}

\begin{table}[htbp]
	\vspace{-1em}
	\caption{The Network Setting}
	\begin{center}
		\begin{tabular}{c l c}
			\hline
			\textbf{Symbol}&\textbf{Definition} & \textbf{Value} \\
			\hline
			$D$ & Network scale   & 800m \\
			$d_0$& Near field threshold   & 10m \\
			$d_1$ & Far field threshold  & 50m \\
			$P$ & Transmit power  & 1W \\
			$N_0$ & Noise power   & $1\times 10^{-12}$ W \\
			$M$& Number of clusters   & 25 \\
			$N/\textrm{rank}(\mbB_m)$& Spike ratio   & $0.7$ \\
			$\beta$&\begin{tabular}[c]{@{}l@{}}
				The ratio between the number\\ of users and the number of BSs\tablefootnote{$\beta<1$ corresponds  to the case where the number of users is smaller than the number of BSs, and $\beta>1$ corresponds to the case where the number of users is larger than the number of BSs. Here we randomly choose one value in each aforementioned case, in order to show the generality of our algorithm.}
			\end{tabular} & 0.5, 8\\
			\hline
		\end{tabular}
		\label{parameters}
	\end{center}
	\vspace{-1em}
\end{table}

To illustrate the high accuracy of our  TOSE algorithm, we compute the values of $\widehat{C}_m$ obtained by TOSE  and the conventional method based on the Cholesky decomposition, respectively, and plot the results under different scenarios in Fig. \ref{tose_results}. Note that we choose the method based on Cholesky decomposition as the baseline to derive the accurate values of $\widehat{C}_m$ according to \eqref{cap2}, and use our TOSE algorithm to estimate $\widehat{C}_m$. It can be observed from Fig. \ref{tose_results} that our TOSE algorithm can accurately estimate $\widehat{C}_m$ under different network settings. The estimation errors are indeed less than 5\%, which means TOSE is sufficiently accurate in practice.

The computational time of TOSE and the Cholesky decomposition based method are plotted in Fig. \ref{tose_time}. 
By data fitting, we can derive the empirical complexity of TOSE is $O(J_m)$, 
while that of the Cholesky decomposition based method is around $O(J_m^3)$. 
This results indicates the overwhelming superiority in computational cost of TOSE. Combining Fig. \ref{tose_results} and Fig. \ref{tose_time}, we can find that TOSE has the same accuracy as the Cholesky decomposition based method, but has much lower computational time.

In Fig. \ref{real_appro}, we show the comparisons between the theoretical values of the capacity  $C_m$ in \eqref{cap1} and the  approximation of   $\widehat C_m$ in \eqref{cap2} obtained by TOSE,  varying the number of clusters. We set the number of BSs in each cluster around 100, so increasing the number of clusters is equivalent to increasing the density of network nodes, under a fixed network area. It can be observed from Fig. \ref{real_appro}  that as the number of clusters increases, the approximated capacity $\widehat C_m$ gets closer and closer to the theoretical capacity $ C_m$, 
which means that  our proposed TOSE algorithm has a higher accuracy on the capacity estimation  for the ultra-dense  networks.

\section{Conclusion}

Capacity is the most important performance metric of wireless networks.  
Determining the capacity of ultra-dense networks in a fast and accurate  way is challenging.
In this paper, we propose a TOSE algorithm to estimate the average cluster capacity,  which is accurate, fast, and general. 
Through making use of the spiked model in RMT, we realize a fast approximation of the top $N$ eigenvalues of a large-dimensional random matrix. The capacity can thus be estimated, avoiding the complex steps to calculate the exact eigenvalues.
Both analytical and simulated results show that, our TOSE algorithm has a linear time complexity, which is much lower than the polynomial time of exiting eigenvalue-calculation-based methods.
Simulation results also show that the accuracy of TOSE is almost the as the exiting eigenvalue calculation based method, on capacity estimation. The estimation error is below 5\%. 
More importantly, TOSE has superior generality, since it is independent of the BS distribution, the user distribution, and the shape of the network area.

\section*{Acknowledgement}

We would like to thank Prof. Hao  Wu for helpful discussions and valuable comments in improving the quality of the manuscript.

\normalem
\bibliographystyle{IEEEtran}

\end{document}